\documentclass[%
showpacs,preprint,
amsmath,amssymb,
aps,
]{revtex4-1}
\usepackage{color}
\usepackage{array,multirow} 
\usepackage{graphicx}
\usepackage{dcolumn}
\usepackage{bm}

\usepackage{mathpazo}
\usepackage{amsmath}
\usepackage{amssymb}
\usepackage{cancel}
\usepackage{graphicx}

\usepackage{ae}            
\usepackage{amsfonts}
\usepackage{makeidx}
\usepackage{lmodern}
\usepackage{colortbl}
\usepackage{indentfirst}
\usepackage{multirow}
\usepackage{array}
\usepackage{caption}
\usepackage[subrefformat=parens,labelformat=parens]{subcaption}
\usepackage{mathtools}

\usepackage{url,hyperref}

\hypersetup{colorlinks,allcolors=blue}

\providecommand{\tabularnewline}{\\}

\usepackage{amsthm}

\usepackage[english]{babel}
\usepackage{adjustbox}

\providecommand{\tabularnewline}{\\}
\providecommand{\ZZ}{\mathbb{Z}}

\usepackage{babel}

\makeatother

\usepackage{babel}
\definecolor{darkred}{RGB}{139, 0, 0}

\begin{document}

	\title{Axion dark matter in a $3-3-1$ model}

	\author{J. C. Montero}
	\email{montero@ift.unesp.br}
	\affiliation{Universidade Estadual Paulista (UNESP),  Instituto de F\'isica Te\'orica (IFT), S\~ao Paulo. R. Dr. Bento Teobaldo Ferraz 271, Barra Funda, S\~ao Paulo - SP, 01140-070, Brasil}
	
		\author{Ana R. Romero Castellanos}
	\email{arromero@ifi.unicamp.br}
	\affiliation{Instituto de F\'isica Gleb Wataghin - UNICAMP, 13083-859, Campinas,
		SP, Brasil.}
	
	\author{B. L. S\'anchez-Vega}
	\email{bruce.sanchez@ufabc.edu.br}
	\affiliation{ Universidade Federal do ABC (UFABC), Centro de Ci\^encias Naturais
		e Humanas,\\
		Av. dos Estados, 5001, 09210-580, Santo Andr\'e, SP, Brasil}
	\affiliation{Instituto de F\'isica Gleb Wataghin - UNICAMP, 13083-859, Campinas,
		SP, Brasil.}

  \begin{abstract}
  Slightly extending a right-handed neutrino version of the $3-3-1$ model, we show that it is not only possible to solve the strong $CP$ problem but also to give the total dark matter abundance reported by the Planck collaboration. Specifically, we consider the possibility of introducing a $3-3-1$ scalar singlet to implement a gravity stable Peccei-Quinn mechanism in this model. Remarkably, for allowed regions of the parameter space, the arising axions with masses $m_a\approx$ meV can both make up the total dark matter relic density through nonthermal production mechanisms and be very close to the region to be explored by the IAXO helioscope.

  \end{abstract}

\maketitle

\section{Introduction\label{introduction}}

The impressive observation that almost thirty percent of the energy
content of the Universe is due to dark matter (DM) is challenging
our understanding of particle physics and cosmology. For a historical review see Ref.~\cite{Bertone:2016nfn}. Much effort
have been done in order to unravel the nature of DM. Experiments designed
to detect weakly interacting massive particles (WIMPs), the, so far, DM candidate paradigm, have failed in
providing positive results~\cite{BERTONE2005279, ANDP201200116}. At the same time, the Large Hadron Collider (LHC) has not been able
to produce any signal of a DM candidate, as is the case of the lightest
supersymmetric  partners of the standard model (SM) neutral
particles (gauge or scalar), called neutralinos, or gravitinos (partners
of the graviton)~\cite{Aaboud2016}.

As a consequence of these negative results, it is noticeable the growing
interest in studying axions and axionlike particles (ALPs) because they
are well motivated alternatives to WIMPs. Moreover, they can be linked
to solutions of still intriguing astrophysical phenomena~\cite{Dias:2014osa}:
(i) ALPs may be the explanation to the TeV photon cosmic transparency
if there are gamma ray $\longleftrightarrow$ ALP oscillations. If
so, gamma rays could be converted to ALPs due to the magnetic fields
near active galactic nuclei, for instance, traveling ``freely'' for a long
distance to our galaxy and then reconverted into gamma rays in the
galactic magnetic fields; (ii) also, ALPs may explain the anomalous
energy loss of white dwarfs because from the luminosity of this kind
of stars it is inferred that a new energy loss mechanism is needed.
In the present scenario, this mechanism could be related to axions
or ALPs bremsstrahlung if they  directly couple to electrons. All these
astrophysical processes constrain the relevant parameters
describing axions and ALPs physics. In fact, besides these theoretical
arguments for considering axions and/or ALPs, there is also much experimental
effort searching for this kind of particles~\cite{experiencias}. A variety of experiments
have been designed and, in general, they are classified as haloscopes,
helioscopes and light-shining through a wall, and most of them are
based on the conversion of axions or ALPs into gamma rays in the presence
of strong magnetic fields~\cite{Battaglieri:2017aum}.

The axion field was initially introduced as a dynamical solution for the so-called strong CP problem. This problem comes from the extra term which has to be added to the QCD Lagrangian due to the nontrivial structure of the QCD vacuum:
$$
\mathcal{L}_{\theta}=\theta\frac{g_s^2}{32\pi^2}G^{a\,\mu\nu}
{\tilde {G}}^{a}_{\mu\nu},$$
where $G^{a\,\mu\nu}$ is the gluon field strength and ${\tilde {G}}^{a}_{\mu\nu}$ its dual.
This $\theta$--term violates P, T and $CP$ symmetries and, hence, it induces a neutron electric dipole moment (NEDM). 
In order to be in agreement with experimental NEDM data the value of the $\theta$ parameter must be 
$\theta \lesssim 0.7\times10^{-11}$~\cite{kim2010axions}. The strong $CP$ problem is, then, to explain why this parameter is so small. 
After including weak interactions, the coefficient of the $G{\tilde{G}}$ term changes to $\bar{\theta} = \theta - \arg \det\,M_q$, 
where $M_q$ is the quark mass matrix. The Peccei--Quinn (PQ) solution to this problem is implemented by introducing a global U$(1)$ 
symmetry that must be spontaneously broken and afflicted by a color anomaly. The axion is then the Nambu--Goldstone boson associated to 
the breaking of that U$(1)$ symmetry, which is now known as the U$(1)_{\textrm{PQ}}$ symmetry. After including the axion field, $a(x),$ 
the total Lagrangian has a term proportional to the color anomaly $N_{\textrm{C}}$:
$$
\mathcal{L}_{\textrm{Total}}=\mathcal{L}_{\textrm{SM}}+
{\bar{\theta}}\frac{g_s^2}{32\pi^2}G^{a\,\mu\nu}
{\tilde {G}}^{a}_{\mu\nu}+\frac{a(x)}{\tilde{f}_a/N_{\textrm{C}}}\frac{g_s^2}{32\pi^2}G^{a\,\mu\nu}
{\tilde {G}}^{a}_{\mu\nu}+\textrm{kinetic}+\textrm{interactions},$$
where $\tilde{f}_a/N_{\textrm{C}}\equiv f_a$ is the axion-decay constant and it is related to the magnitude of the vacuum expectation value (VEV) that breaks the  U$(1)_{\textrm{PQ}}$ symmetry. We also have that the divergence of the PQ current, $\partial_\mu J^\mu_{\textrm{PQ}}$, is $N_{\textrm{C}}\frac{g_s^2}{32\pi^2}G^{a\,\mu\nu}{\tilde {G}}^{a}_{\mu\nu} \neq 0$. 
Hence, the $CP$ violating term $G{\tilde{G}}$ is now proportional to
(${\bar{\theta}}+N_{\textrm{C}}\,a(x)/\tilde{f}_a)$ and it is shown that $\langle a(x) \rangle = - {\tilde{f}_a \bar{\theta}}/{N_{\textrm{C}}}$ 
minimizes the axion effective potential so that, when the axion field is redefined, $a(x)\rightarrow a(x) - \langle a(x)\rangle,$ the $CP$ 
violating term $G{\tilde{G}}$ is no longer present in the Lagrangian, solving in this way the strong $CP$ problem. Although the axion is massless 
at tree level, it is, in fact, a pseudo-Nambu-Goldstone boson since it gains a  mass due to nonperturbative QCD effects related to the 
$\textrm{U}(1)_{\textrm{PQ}}$ color anomaly. The axion mass and all its couplings are governed by the value of $f_a$. 
The original conception of the axion was ruled out long ago because $f_a$ was thought to be near the electroweak scale, implying in a
``visible'' axion, in contradiction with laboratory and astrophysical constraints. Few years after the PQ proposal it was realized that for 
large enough values of $f_a$ the axion could be a cold dark matter candidate~\cite{PRESKILL1983127,ABBOTT1983133,DINE1983137}. In fact,
for high symmetry breaking scales, the axion is a nonbaryonic extremely weakly-interacting massive particle, stable on cosmological time scales, which makes it a candidate to dark matter. Later in the text we discuss the constraints on $f_a$ coming from NEDM, ``invisibility'' of the axion, and astrophysical data.

In order to consider the axion  a viable DM candidate we must  deal with its relic abundance which strongly depends on the history of the Universe. In particular, the cosmological scenario for the axion production changes significantly  if the PQ symmetry is broken before or after the inflationary expansion of the Universe.  
The main issue related to the order of these events concerns the axion-production mechanisms. There are production mechanisms due to topological defects, like axionic strings and domain walls, that are comparable to the vacuum misalignment one. Hence, on one hand, if the PQ-symmetry breaking occurs before inflation, inflation will erase  these topological defects. On the other hand, if the PQ-symmetry breaking happens after inflation, it is expected  an additional number of axions to be produced due to the decay of the topological defects, affecting directly the relic abundance estimative. In this work we consider  axions as DM candidates in the so-called post-inflationary scenario, when the reheating temperature, $T_R$, is high enough to restore the PQ symmetry, $T_R>T_C \sim f_a$, which will be broken at a later time, when the temperature of the Universe falls below the critical temperature $T_C$.

As we can see, axions present some features with relevant implications not only in particle physics but also in cosmology and it is also a strong indication that physics beyond the SM is in order. In this vein a large variety of models, extensions
of the SM, has been proposed. Most of them claim for very appealing
achievements relating the DM solution to another yet unsolved
issue in particle physics~\cite{Carvajal:2017gjj,Sanchez-Vega:2015qva,Sanchez-Vega:2014rka}, as it is the case of the lightness of the
active neutrino masses, the smallness of the strong $CP$ violation,
or the hierarchy problem, for instance.

Among others, a way of introducing new physics is to consider a model
with a larger symmetry group. In particular, there is a class of models
based on the $\textrm{SU}(3)_{C}\otimes \textrm{SU}(3)_{L}\otimes \textrm{U}(1)_{X}$ gauge group
(the so called $3-3-1$ models, for shortness), which are interesting
extensions of the SM. In general, these $3-3-1$ models bring welcome
features which we review very shortly here. We can take advantage
of the larger group representation to choose the matter content in
order to introduce new degrees of freedom which are appropriate to
implement, for instance, a mechanism to generate tiny active neutrino
masses, in the lepton sector. The quark sector will also have new
degrees of freedom and, depending on the particular representation,
the model can have quarks with exotic electric charges or not. The
issue of the chiral anomaly cancellation is solved provided we have
the same number of triplets and anti-triplets, including color counting.
Then, considering that we have the same number of lepton and quark
families, say $n_{f}$, we find that $n_{f}$ must be three or a multiple
of three. However, from the QCD asymptotic freedom we find that the
number of families must be just three in order to get the correct,
negative, sign of the renormalization group $\beta$ function. Note
that, contrarily to the SM, the total number of families must be considered
altogether in order to get the model anomaly free. Hence, the number
of families and the number of colors are related to each other by
the anomaly cancellation condition. This fact is a direct consequence
of the $3-3-1$ gauge invariance and it can be seen as a hint to the
solution to the family replication issue. We can still mention other
interesting features: (i) the electric charge quantization does not depend
if neutrinos are Majorana or Dirac fermions~\cite{Pires:1998}; (ii) the model
described in Refs.~\cite{Pleitez_etal_1992,Frampton_1992,Z2_331_3} presents the relation $t^{2}=(g^{\prime}/g)^{2}=\sin^{2}\theta_{W}/(1-4\sin^{2}\theta_{W})$,
which relates the $\textrm{U}(1)_{X}$ and the $\textrm{SU}(3)_{L}$ coupling constants,
$g^{\prime}$ and $g$, respectively, to the electroweak $\theta_{W}$
angle. This relations shows a Landau-like pole at some ${\cal O}(\textrm{TeV})$,
energy scale, $\mu$, for which $\sin^{2}\theta_{W}(\mu)=1/4$~\cite{Dias2005},
and it would be an explanation to the observed value $\sin^{2}\theta_{W}(M_{Z})<1/4$.
(iii) The Peccei-Quinn symmetry, usually introduced to solve the strong
$CP$ problem, can be introduced in a natural way~\cite{Montero2011_PQ}.
In this work we consider a version of a $3-3-1$ model
where a gravity stable PQ mechanism can be implemented.  We analyze the conditions under which the axion, resulting from the spontaneous breaking of the PQ symmetry in this model, can be considered a dark matter candidate. 

This work is organized as follows. In  Sec. \ref{model}, we present the general features of the $3-3-1$ model, including its matter content, 
Yukawa interactions and scalar potential. In Sec.~\ref{PQmechanism} we show the main steps to make the axion invisible and the PQ mechanism stable 
against gravitational effects. We  also show the axion effective potential from which   its mass is derived. In Sec.~\ref{axiondarkmatter} we 
consider the axion production mechanisms in order to compute its abundance in the Universe. Results for the vacuum misalignment and  decay of the 
string and string-wall system mechanisms are given. In Sec.~\ref{results} we confront the predictions from the previous section with the 
observational constraints, coming mainly from the Planck-collaboration results for the DM abundance, the NEDM data and direct axion searches, 
in order to constrain the parameter space of the model.  Section \ref{conclusions} is devoted to our final discussions and conclusions.

\section{Briefly reviewing the model\label{model}}

We consider the $3-3-1$ model with right-handed neutrinos,
$N_{a}$, in the same multiplet as the SM leptons, $\nu_{a}$ and
$e_{a}$. In other words, in this model all of the left-handed leptons,
$F_{aL}=\left(\nu_{a},\,e_{a},\,N_{a}^{c}\right)_{L}^{\textrm{T}}$
with $a=1,2,3$, belong to the same $\left(1,\,\mathbf{3},\,-1/3\right)$
representation, where the numbers inside the parenthesis denote the
quantum numbers of $\textrm{SU}(3)_{C}$, $\textrm{SU}(3)_{L}$ and $\textrm{U}(1)_{X}$ gauge
groups, respectively. This model was proposed in Refs.~\cite{neutrino_masses,Montero:1993}
and it has been subsequently considered in Refs.~\cite{Pires_daSilva_331,Dias_Pires_daSilva_331,Z2_331_2,Dong:2007,Mizukoshi_Pires_Queiroz_daSilva_331,PhysRevD.69.077702,Montero2011_PQ,331_Farinaldo,Montero:2015,Avelino}.
It shares appealing features with other versions of $3-3-1$ models~\cite{Pleitez_etal_1992,Frampton_1992,Z2_331_3,Clavelli_etal_1974,Lee_Weinberg_1977,Lee_Shrock_1978,Singer_1979,Singer_etal_1980}.
Furthermore, the existence of right-handed neutrinos allows mass terms at tree level, but it is necessary to go to the one-loop level to obtain neutrino masses in agreement with experiments~\cite{Dong:2007}.

The remaining left-handed fermionic fields of the model belong to
the following representations 
\begin{align}
\textrm{Quarks: }Q_{L} & =\left(u_{1},\,d_{1},\,u{}_{4}\right)_{L}^{\textrm{T}}\sim\left(\mathbf{3},\,\mathbf{3},\,1/3\right),\\
Q_{bL} & =\left(d_{b},\,u_{b},\,d_{b+2}\right)_{L}^{\textrm{T}}\sim\left(\mathbf{3},\,\bar{\mathbf{3}},\,0\right),
\end{align}
where $b=2,\,3$; and ``$\sim$'' means the transformation properties
under the local symmetry group. Additionally, in the right-handed
field sector we have 
\begin{align}\label{eq:rep-leptons}
\textrm{Leptons: }e_{aR} & \sim\left(1,\,1,\,-1\right),\\
\textrm{Quarks:\,\ }u_{sR} & \sim\left(\mathbf{3},\,1,\,2/3\right),\quad d_{tR}\sim\left(\mathbf{3},\,1,\,-1/3\right),
\end{align}
where $a=1,2,3$; $s=1,\dots,4$ and $t=1,\dots,5$.

In order to generate the fermion and boson masses, the $\textrm{SU}(3)_{C}\otimes\textrm{SU}(3)_{L}\otimes\textrm{U}(1)_{X}$
symmetry must be spontaneously broken to the electromagnetic group,
i.e., to the $\textrm{U}\left(1\right)_{Q}$ symmetry, where $Q$ is the electric charge.  To do this, it is necessary to introduce, at least, three SU$\left(3\right)_{L}$ triplets, $\eta,\,\rho,\,\chi$, as shown in Ref.~\cite{Montero:2015}, which are given by 
\begin{equation}
\eta=\left(\eta_{1}^{0},\,\eta_{2}^{-},\,\eta_{3}^{0}\right)^{\textrm{T}}\sim\left(1,\:\mathbf{3},\,-1/3\right),\quad\rho=\left(\rho_{1}^{+},\,\rho_{2}^{0},\,\rho_{3}^{+}\right)^{\textrm{T}}\sim\left(1,\,\mathbf{3},\,2/3\right),
\end{equation}
\begin{equation}
\quad\chi=\left(\chi_{1}^{0},\,\chi_{2}^{-},\,\chi_{3}^{0}\right)^{\textrm{T}}\sim\left(1,\:\mathbf{3},\,-1/3\right).
\end{equation}

Once these fermionic and bosonic fields are introduced in the model, we can write the most general Yukawa Lagrangian, invariant under the local gauge group, as follows 
\begin{equation}
\mathcal{L}_{\textrm{Yuk}}=\mathcal{L}_{\textrm{Yuk}}^{\rho}+\mathcal{L}_{\textrm{Yuk}}^{\eta}+\mathcal{L}_{\textrm{Yuk}}^{\chi},\label{eq:yukawa}
\end{equation}
with 
\begin{eqnarray}
\mathcal{L}_{\textrm{Yuk}}^{\rho} & = & \alpha_{t}\bar{Q}_{L}d_{tR}\rho+\alpha_{bs}\bar{Q}_{bL}u_{sR}\rho^{*}+\text{Y}_{aa^\prime}\epsilon_{ijk}\left(\bar{F}_{aL}\right)_{i}\left(F_{a^\prime L}\right)_{j}^{c}\left(\rho^{*}\right)_{k}+\textrm{Y}^\prime_{aa^\prime}\bar{F}_{aL}e_{a^\prime R}\rho\nonumber \\
 &  & +\textrm{H.c.,}\label{5}\\
\mathcal{L}_{\textrm{Yuk}}^{\eta} & = & \beta_{s}\bar{Q}_{L}u_{sR}\eta+\beta{}_{bt}\bar{Q}_{bL}d_{tR}\eta^{*}+\textrm{H.c.},\label{6}\\
\mathcal{L}_{\textrm{Yuk}}^{\chi} & = & \gamma_{s}\bar{Q}_{L}u_{sR}\chi+\gamma{}_{bt}\bar{Q}_{bL}d_{tR}\chi^{*}+\textrm{H.c.},\label{7}
\end{eqnarray}
where $\epsilon_{ijk}$ is the Levi-Civita symbol and $a^\prime,i,j,k=1,2,3$ and $a$, $b$, $s$, $t$ are in the same range as in Eq.~(\ref{eq:rep-leptons}). It is also straightforward to write down the most general scalar potential consistent with gauge invariance and renormalizability as 
\begin{eqnarray}\label{eq:PotencialesV}
V\left(\eta,\rho,\chi\right) & = & V_{\mathbb{Z}_{2}}\left(\eta,\rho,\chi\right)+V_{\cancel{\mathbb{Z}_{2}}}\left(\eta,\rho,\chi\right);\label{scalarpotential}\\
 \textrm{with }\nonumber \\
V_{\mathbb{Z}_{2}}\left(\eta,\rho,\chi\right) & = & -\mu_{1}^{2}\eta^{\dagger}\eta-\mu_{2}^{2}\rho^{\dagger}\rho-\mu_{3}^{2}\chi^{\dagger}\chi\nonumber \\
 &  & +\lambda_{1}\left(\eta^{\dagger}\eta\right)^{2}+\lambda_{2}\left(\rho^{\dagger}\rho\right)^{2}+\lambda_{3}\left(\chi^{\dagger}\chi\right)^{2}+\lambda_{4}\left(\chi^{\dagger}\chi\right)\left(\eta^{\dagger}\eta\right)\nonumber \\
 &  & +\lambda_{5}\left(\chi^{\dagger}\chi\right)\left(\rho^{\dagger}\rho\right)+\lambda_{6}\left(\eta^{\dagger}\eta\right)\left(\rho^{\dagger}\rho\right)+\lambda_{7}\left(\chi^{\dagger}\eta\right)\left(\eta^{\dagger}\chi\right)\nonumber \\
 &  & +\lambda_{8}\left(\chi^{\dagger}\rho\right)\left(\rho^{\dagger}\chi\right)+\lambda_{9}\left(\eta^{\dagger}\rho\right)\left(\rho^{\dagger}\eta\right)+[\lambda_{10}\left(\chi^{\dagger}\eta\right)^{2}+\textrm{H.c.}];\label{vz2}\\
V_{\cancel{\mathbb{Z}_{2}}}\left(\eta,\rho,\chi\right) & = & -\mu_{4}^{2}\chi^{\dagger}\eta\nonumber \\
 &  & +\lambda_{11}\left(\chi^{\dagger}\eta\right)\left(\eta^{\dagger}\eta\right)+\lambda_{12}\left(\chi^{\dagger}\eta\right)\left(\chi^{\dagger}\chi\right)+\lambda_{13}\left(\chi^{\dagger}\eta\right)\left(\rho^{\dagger}\rho\right)\nonumber \\
 &  & +\lambda_{14}\left(\chi^{\dagger}\rho\right)\left(\rho^{\dagger}\eta\right)+\lambda_{15}\epsilon_{ijk}\eta_{i}\rho_{j}\chi_{k}+\textrm{H.c.}
\end{eqnarray}
We have divided the total scalar potential $V\left(\eta,\rho,\chi\right)$
in two pieces, $V_{\mathbb{Z}_{2}}\left(\eta,\rho,\chi\right)$, invariant under the $\ZZ_{2}$ discrete
symmetry ($\chi\rightarrow-\chi$, $u_{4R}\rightarrow-u{}_{4R}$,
$\,d{}_{\left(4,5\right)R}\rightarrow-d{}_{\left(4,5\right)R}$, and all the other fields even by the symmetry), and 
$V_{\cancel{\mathbb{Z}_2}}\left(\eta,\rho,\chi\right)$, which breaks $\mathbb{Z}_2$.  This discrete symmetry is motivated
by the implementation of the PQ mechanism as shown
below.

It is well known that the minimal vacuum structure needed to give
masses to all the particles in the model is 
\begin{equation}\label{eq14}
\left\langle \rho\right\rangle =\frac{1}{\sqrt{2}}\left(0,\,v_{\rho_{2}^{0}},\,0\right)^{\textrm{T}},\,\,\left\langle \eta\right\rangle =\frac{1}{\sqrt{2}}\left(v_{\eta_{1}^{0}},\,0,\,0\right)^{\textrm{T}},\,\,\left\langle \chi\right\rangle =\frac{1}{\sqrt{2}}\left(0,\,0,\,v_{\chi_{3}^{0}}\right)^{\textrm{T}},
\end{equation}
which correctly reduces the $\textrm{SU}\left(3\right)_{C}\otimes\textrm{SU}\left(3\right)_{L}\otimes\textrm{U}\left(1\right)_{X}$
symmetry to the $\textrm{U}\left(1\right)_{Q}$ one. In principle,
the remaining neutral scalars, $\eta_{3}^{0}$ and $\chi_{1}^{0}$,
can also gain VEVs. However, in this case, dangerous Nambu-Goldstone bosons can arise in the physical spectrum, as shown in Ref.~\cite{Sanchez-Vega:2016dwe}. In this
paper, we are going to consider only the minimal vacuum structure
given in Eq.~(\ref{eq14}).

\section{Implementing a gravity stable PQ mechanism}\label{PQmechanism}

The key ingredient to implement the PQ mechanism is the invariance
of the entire Lagrangian under a global $\textrm{U}\left(1\right)$
symmetry, called $\textrm{U}\left(1\right)_{\text{PQ}}$, which must be both afflicted by a color anomaly and spontaneously broken~\cite{Kim1979,Dine1981,Georgi1982,Srednicki:1985xd}.
In general, the implementation of the PQ mechanism in the $3-3-1$
models is relatively straightforward~\cite{PhysRevD.69.077702,Montero2011_PQ}.
In particular, in Ref.~\cite{Montero2011_PQ} a gravitationally stable
PQ mechanism for the model considered here is successfully implemented.
We are going to review its main results for completeness.

First of all, we search for all $\textrm{U}\left(1\right)$ symmetries
of the Lagrangian given in Eqs.~(\ref{eq:yukawa}) and (\ref{scalarpotential}).
Doing so, we find only two symmetries, U$\left(1\right)_{X}$ and
U$\left(1\right)_{B}$, which clearly do not satisfy the two minimal conditions required 
for the $\textrm{U}\left(1\right)_{\text{PQ}}$ symmetry. See Table \ref{table 1} for the quantum number assignments of the fields for these symmetries. In other words,
the U$(1)_{\text{PQ}}$ is not naturally allowed by the gauge symmetry.
\begin{table}[th]
\caption{The U$\left(1\right)$ symmetries of the Lagrangian given by
Eqs.~(\ref{eq:yukawa}) and (\ref{scalarpotential}).}
\label{table 1}\centering 
\begin{tabular}{ccccccccc}
\hline 
 & \,\, $Q_{L}$  & \,\, $Q_{iL}$  & \,\, ($u_{aR}$, $u_{4R}$)  & \,\, ($d_{aR}$, $d_{\left(4,5\right)R}$)  & \,\,$F_{aL}$  & \,\, $e_{aR}$  & \,\, $\rho$  & \,\, ($\chi$, $\eta$) \tabularnewline
\hline 
U$\left(1\right)_{X}$  & $1/3$  & $0$  & $2/3$  & $-1/3$  & $-1/3$  & $-1$  & $2/3$  & $-1/3$ \tabularnewline
U$\left(1\right)_{B}$  & $1/3$  & $1/3$  & $1/3$  & $1/3$  & $0$  & $0$  & $0$  & $0$ \tabularnewline
\hline 
\end{tabular}
\end{table}
However, if the Lagrangian is slightly modified by imposing a $\ZZ_{2}$
discrete symmetry such that $\chi\rightarrow-\chi$, $u_{4R}\rightarrow-u{}_{4R}$,
$\,d{}_{\left(4,5\right)R}\rightarrow-d{}_{\left(4,5\right)R}$, all
terms in $V_{\cancel{\mathbb{Z}_{2}}}\left(\eta,\rho,\chi\right)$
are forbidden. In addition, the Yukawa Lagrangian interactions given
in Eqs. (\ref{5}-\ref{7}) are slightly modified to 
\begin{eqnarray}
\mathcal{L}_{\textrm{Yuk}}^{\rho} & = & \alpha_{a}\bar{Q}_{L}d_{aR}\rho+\alpha_{ba}\bar{Q}_{bL}u_{aR}\rho^{*}+\textrm{Y}_{aa^\prime}\varepsilon_{ijk}\left(\bar{F}_{aL}\right)_{i}\left(F_{bL}\right)_{j}^{c}\left(\rho^{*}\right)_{k}+\textrm{Y}^\prime_{aa^\prime}\bar{F}_{aL}e_{a^\prime R}\rho+\nonumber \\
 &  & \textrm{H.c.,}\label{10}\\
\mathcal{L}_{\textrm{Yuk}}^{\eta} & = & \beta_{a}\bar{Q}_{L}u_{aR}\eta+\beta{}_{ba}\bar{Q}_{bL}d_{aR}\eta^{*}+\textrm{H.c.},\label{11}\\
\mathcal{L}_{\textrm{Yuk}}^{\chi} & = & \gamma_{4}\bar{Q}_{L}u_{4R}\chi+\gamma{}_{b\left(b+2\right)}\bar{Q}_{bL}d_{\left(b+2\right)R}\chi^{*}+\textrm{H.c.}\,.\label{12}
\end{eqnarray}
Consequently, with the imposition of this $\ZZ_{2}$ symmetry a U$\left(1\right)_{\text{PQ}}$ symmetry is automatically introduced with the charges given in Table \ref{table 2}. 

\begin{table}[th]
\caption{The U$\left(1\right)_{\text{PQ}}$ charges in the model with a $\ZZ_{2}$
discrete symmetry such that $\chi\rightarrow-\chi$, $u_{4R}\rightarrow-u{}_{4R}$,
and $\,d{}_{\left(4,5\right)R}\rightarrow-d{}_{\left(4,5\right)R}$
.}\label{table 2}\centering %
\begin{tabular}{ccccccccc}
\hline 
 & \,\, $Q_{L}$  & \,\, $Q_{iL}$  & \,\, ($u_{aR}$, $u_{4R}$)  & \,\, ($d_{aR}$, $d_{\left(4,5\right)R}$)  & \,\,$F_{aL}$  & \,\, $e_{aR}$  & \,\, $\rho$  & \,\, ($\chi$, $\eta$) \tabularnewline
\hline 
U$\left(1\right)_{\text{PQ}}$  & $-2$  & $2$  & $0$  & $0$  & $1$  & $3$  & $-2$  & $-2$ \tabularnewline
\hline 
\end{tabular}
\end{table}

As $\eta,\rho,\chi$ get VEVs, an axion appears in the physical spectrum.
However, it is a visible axion because the U$\left(1\right)_{\text{PQ}}$
symmetry is actually broken by $v_{\rho_{2}^{0}}$, which is upper
bounded by the value of $v_{\rm{SM }}\simeq246$ GeV, as shown in
Refs.~\cite{Montero2011_PQ,Sanchez-Vega:2016dwe}. Hence, this scenario
is ruled out~\cite{Bardeen1987}. Nevertheless, a singlet scalar,
$\phi\sim (1,1,0)$, can be introduced in order to make the axion invisible. Its
role is to break the PQ symmetry at an energy scale much larger than the electroweak one. This field does not couple directly to quarks and
leptons, however it couples to the scalar triplets, $\eta$, $\rho$ and $\chi$, through Hermitian terms and the non-Hermitian term $\lambda_{\text{PQ}}\epsilon^{ijk}\eta_{i}\rho_{j}\chi_{k}\phi$,  from which it gets a PQ charge equal to $6$, cf.  Table \ref{table 2}. Notice that this term is allowed as long as the $\phi$ field
is odd under the $\ZZ_{2}$ symmetry, i.e., $\ZZ_{2}\left(\phi\right)=-\phi$.

Although the $\ZZ_{2}$ discrete symmetry apparently introduces the
PQ mechanism in the model, there are two issues with it. First,
the $\ZZ_{2}$ and gauge symmetries allow some renormalizable terms in the scalar potential,
such as $\phi^{2}$, $\phi^{4}$, $\rho^{\dagger}\rho\phi^{2}$, $\eta^{\dagger}\eta\phi^{2}$,
$\chi^{\dagger}\chi\phi^{2}$, that explicitly
violate the PQ symmetry in an order low enough to make the PQ mechanism
ineffective. Second, since the PQ symmetry is global, it is expected to be broken by gravitational effects \cite{kamionkowski1992planck,holman1992solutions}. Thus, a mechanism to stabilize the axion solution has to be introduced. As usual, the entire Lagrangian
is considered to be invariant under a $\ZZ_{D}$ discrete gauge symmetry
(anomaly free)~\cite{PhysRevLett.62.1221,Ibanez1993301,Babu2003,Babu2003322,PhysRevD.69.077702,Montero2011_PQ}
and, in addition, this symmetry is supposed to induce the U$\left(1\right)_{\text{PQ}}$
symmetry. For $\ZZ_{D\geq10}$ it is found that all
effective operators of the form $\phi^{N}/M_{\text{Pl}}^{N-4}$ (where
$N\geq D$ is a positive integer and $M_{\textrm{Pl}}$
is the reduced Planck mass) that can jeopardize the PQ mechanism are suppressed.
In particular, in Ref. \cite{Montero2011_PQ}  two different symmetries,
$\ZZ_{10}$ and $\ZZ_{11}$, were found to stabilize the PQ mechanism for the Lagrangian given by  Eqs. (\ref{scalarpotential},\ref{10}-\ref{12}). The specific charge assignments for these
symmetries are shown in Table \ref{table 3}. Note that the term $\lambda_{15}\epsilon_{ijk}\eta_{i}\rho_{j}\chi_{k}$ in the scalar potential is prohibited by both of these discrete symmetries and it must be removed from the entire Lagrangian. 

\begin{table}[th]
\caption{The charge assignment for $\ZZ_{D}$ that stabilizes the PQ mechanism
in the considered $3-3-1$ model.}

\label{table 3}\centering %
\begin{tabular}{cccccccccc}
\hline 
 & \,\, $Q_{L}$  & \,\, $Q_{iL}$  & \,\, ($u_{aR}$, $u_{4R}$)  & \,\, ($d_{aR}$, $d_{\left(4,5\right)R}$)  & \,\,$F_{aL}$  & \,\, $e_{aR}$  & \,\, $\rho$  & \,\, ($\chi$, $\eta$)  & $\ \phi$ \ \tabularnewline
\hline 
$\ZZ_{10}$  & $+7$  & $+5$  & $+1$  & $+1$  & $+7$  & $+1$  & $+6$  & $+6$  & $+2$ \tabularnewline
\hline 
$\ZZ_{11}$  & $+7$  & $+6$  & $+1$  & $+1$  & $+8$  & $+2$  & $+6$  & $+6$  & $+4$ \tabularnewline
\hline 
\end{tabular}
\end{table}

We remark that both the $\ZZ_{10}$ and $\ZZ_{11}$ discrete symmetries in Table \ref{table 3}
are anomaly free. This type of discrete symmetry is known as  gauge discrete $\ZZ_{N}$ symmetry and it is assumed to be a remnant
of a gauge (local) symmetry valid at very high energies, \cite{PhysRevLett.62.1221}.
The anomaly-free conditions are necessary in order to truly protect the PQ mechanism against gravity effects \cite{Ibanez1991291,PhysRevD.45.1424,Ibanez1993301,Luhn:2008sa},
Specifically, these discrete symmetries satisfy $A_{3C}(\ZZ_{N})=A_{3L}(\ZZ_{N})=0 \textrm{ Mod}\ N/2$, where $A_{3C}$ and  $A_{3L}$ are the $[\text{SU}(3)_{C}]^{2}\times \ZZ_{N}$, $[\text{SU}(3)_{L}]^{2}\times \ZZ_{N}$ 
anomalies, respectively. Other anomalies, such as $\ZZ_{N}^{3}$, do not give useful low energy constraints because these depend on some arbitrary choices concerning to the full theory.  In particular, the $\ZZ_{N}^{3}$ anomaly depends on the fermions which get masses at very high energy and are integrated out in the low-energy Lagrangian. All the details of these anomaly conditions applied to the $3-3-1$ model can be found in Ref. \cite{Montero2011_PQ}. 

In both cases, the axion, $a\left(x\right)$, is the phase
of the $\phi$ field, i.e., $\phi\left(x\right)\propto\exp\left(i a\left(x\right)/\tilde{f}_{a}\right)$, which implies $\tilde{f}_{a}\approx v_{\phi}$.
As it is well known, to make the axion compatible with astrophysical
and cosmological considerations, the axion-decay constant $f_a$ (related to $\tilde{f}_a$ by $f_{a}=\tilde{f}_a/N_{\textrm{C}}=\tilde{f}_a/N_{DW}$, with $N_{DW}$ being the number of domain walls in the theory. In this model we have $N_{\textrm{C}}=N_{DW}=3$), must be
in the range $10^{9}$ GeV $\lesssim$ $f_{a}$ $\lesssim$
$10^{12}$ GeV (we are assuming a post-inflationary PQ symmetry breaking
scenario). Note that this high value of $f_a=\tilde{f}_{a}/N_{\textrm{C}}\approx v_{\phi}/N_{\textrm{C}}\gg v_{\rho_{2}^{0}},\,v_{\eta_{1}^{0}},\,v_{\chi_{3}^{0}}$,
justifies the approximation in the form of axion eigenstate. It is also important to remember that in this model $v_{\rho_{2}^{0}}^{2}+v_{\eta_{1}^{0}}^{2}=v_{\textrm{SM}}^{2}$
and $v_{\chi_{3}^{0}}$ is expected to be at the TeV energy scale.

Now, we can go further calculating the axion mass, $m_{a}$. In this
model, the axion gains mass because the U$\left(1\right)_{\text{PQ}}$
symmetry is both anomalous under the $\textrm{SU}(3)_{\textrm{C}}$ group and
explicitly broken by gravity-induced operators, $g\phi^{N}/M_{\text{Pl}}^{N-4}$
(with $g=\left|g\right|\exp i\delta$). These operators have a high
dimension ($N\geq10$) because of the protecting $\ZZ_{10}$ or $\ZZ_{11}$
discrete symmetries, as shown in Table~\ref{table 3}. These two effects induce an effective potential
for the axion, $V_{\textrm{eff}}$, from which it is possible to determine the axion mass.\\
In more detail, as the U$\left(1\right)_{\text{PQ}}$ symmetry
is anomalous, we will have a $V_{\text{PQ}}$ term in the effective potential,
which can be written as 
\begin{eqnarray}
V_{\text{PQ}} & = & -m_{\pi}^{2}f_{\pi}^{2}\left[1-\frac{4m_{u}m_{d}}{\left(m_{u}+m_{d}\right)^{2}}\sin^{2}\left(\frac{a\left(x\right)}{2f_{a}}\right)\right]^{1/2},
\label{PotencialPQ}
\end{eqnarray}
where $m_{\pi}\simeq135$ MeV and $f_{\pi}\simeq92$ MeV are the mass
and decay constant of the neutral pion, respectively; $m_u$ and $m_d$ are the masses of the up and down quarks. Note
that $V_{\text{PQ}}$ has a minimum when $\left\langle a\left(x\right)\right\rangle /f_{a}=0$, which solves the strong $CP$ problem in the usual way.

However, because of the PQ symmetry is also explicitly broken by gravity
effects, the effective potential gets another term, $V_{\text{gravity}}$,
which reads 

\begin{eqnarray}\label{gravpotential}
V_{\text{gravity}} & \simeq & -\frac{\left|g\right|v_{\phi}^{N}}{2^{N/2-1}M_{\textrm{Pl}}^{N-4}}\cos\left(\frac{N\,a\left(x\right)}{\tilde{f}_{a}}+\delta_D\right),\label{vgravity}
\end{eqnarray}
where $N=10,11$ for $\ZZ_{10}$ and $\ZZ_{11}$, respectively. The 
phase $\delta_D$ inside the trigonometric function can be written as 
\begin{eqnarray}
\delta_D & = & \delta-N\bar{\theta},\label{fased}
\end{eqnarray}
where $\delta$ is the phase of the $g$ coupling constant 
and $\bar{\theta}$ is the parameter which couples to the gluonic
field strength and its dual. This extra term in the scalar potential,
Eq.~(\ref{vgravity}), has two important consequences. First, it induces
a shift in the value of $\frac{\left\langle a\left(x\right)\right\rangle }{f_{a}}$
where $V_{\textrm{eff}}$ has a minimum. Expanding $V_{\textrm{eff}}=V_{\text{PQ}}+V_{\text{gravity}}$
in powers of $\frac{\left\langle a\left(x\right)\right\rangle }{f_{a}}$,
we find that in the minimum, the axion VEV satisfies
\begin{eqnarray}\label{labelNEDM1}
\left.\frac{|\left\langle a\left(x\right)\right\rangle| }{f_{a}}\right|_{\textrm{min}} & \simeq & \frac{\frac{N\left|g\right|N_{\textrm{DW}}^{N-1}}{2^{\frac{N}{2}-1}}\left(\frac{f_{a}}{M_{\textrm{Pl}}}\right)^{N-2}M_{\textrm{Pl}}^{2}\sin\delta_D}{\frac{m_{\pi}^{2}f_{\pi}^{2}}{f_{a}^{2}}\frac{m_{u}m_{d}}{\left(m_{u}+m_{d}\right)^{2}}+\frac{N^{2}\left|g\right|N_{\textrm{DW}}^{N-2}}{2^{\frac{N}{2}-1}}\left(\frac{f_{a}}{M_{\textrm{Pl}}}\right)^{N-2}M_{\textrm{Pl}}^{2}\cos\delta_D},
\end{eqnarray}
where we have used $v_\phi\approx  \tilde{f}_{a}=N_{\textrm{DW}}f_a.$ Note that for $\left|g\right|=0$ (or for $\delta_D=0$) we have that
$\frac{\left\langle a\left(x\right)\right\rangle }{f_{a}}=0$
in the minimum, as it should be to solve the strong $CP$ problem. However, in the general case, the value of $\frac{\left\langle a\left(x\right)\right\rangle }{f_{a}}$
does not satisfy the NEDM constraint~\cite{kim2010axions}, which imposes
\begin{eqnarray}\label{eq:NEDM}
\frac{\left\langle a\left(x\right)\right\rangle }{f_{a}}=\bar{\theta} & \lesssim & 0.7\times10^{-11}.\label{NEDM}
\end{eqnarray}
In addition, $V_{\text{gravity}}$ brings a mass contribution for
the axion, $m_{a,\textrm{ gravity}}$. From Eq.~(\ref{vgravity})
we obtain 
\begin{eqnarray}\label{eq:mgravity}
m_{a,\textrm{ gravity}}^{2} & = & \frac{N^{2}\left|g\right|N_{\textrm{DW}}^{N-2}}{2^{\frac{N}{2}-1}}\left(\frac{f_{a}}{M_{\textrm{Pl}}}\right)^{N-2}M_{\textrm{Pl}}^{2}\cos\delta_D.
\end{eqnarray}
This contribution can, in general, be much larger than the well-known
axion-mass term coming from the QCD nonperturbative terms, Eq. (\ref{PotencialPQ}), 
\begin{equation}\label{eq:axionQCDmass}
m_{a,\textrm{ QCD}}^{2}=\frac{m_{\pi}^{2}f_{\pi}^{2}}{f_{a}^{2}}\frac{m_{u}m_{d}}{\left(m_{u}+m_{d}\right)^{2}}.
\end{equation}
Thus, in order to maintain the axion mass stable, we are going to
look for values of the parameters $\left|g\right|$, $f_{a}$ and $\delta_D$ for $N=10,11$ that both satisfy the NEDM constraint and leave the axion mass stable $(m_{a,\textrm{ QCD}}\gtrsim m_{a,\textrm{ gravity}})$. 

Before closing this section, it is important to remark that although
the $3-3-1$ model considered in this paper has additional contributions
to $CP$-violating processes that in principle can contribute to the
NEDM, these do not require tuning the model parameters at the same
order of the $\overline{\theta}$ parameter as it was correctly estimated in Ref.~\cite{Montero2011_PQ}. Roughly speaking, the dominant contribution
to the up-quark electric dipole moment, $d_{u}^{e}$, coming from
the interchange of the $\chi$ scalar is of order $\left.d_{u}^{e}\right|_{m_{u}\ll m_{u_{4}},m_{\chi}}\approx\frac{e\left|\gamma_{4}\cdot\gamma{}_{b\left(b+2\right)}\right|\sin\alpha}{48\pi^{2}}\frac{m_{u_{4}}}{m_{\chi}^{2}}{\cal {K}}\left(r\right),$
where $\sin\alpha$ is the sine of the $CP$-violating phase, $\alpha$,
and ${\cal {K}}\left(r\right)=\frac{1}{2r}-\frac{1}{r^{2}}+\frac{1}{r^{3}}\ln\left(1+r\right),$
with $r=\frac{m_{u_{4}}^{2}}{m_{\chi}^{2}}-1$, and where $m_{u}$
is the up-quark mass; $m_{u_{4}}$ and $m_{\chi}$ are the exotic
quark and scalar masses, respectively. For reasonable Yukawa couplings ($\gamma_4$, $\gamma_{b(b+2)}$)
and $CP$-violating phases, and for $m_{u_{4}}$ and $m_{\chi}$ masses
of order of TeV, the $d_{n}^{e}\sim\frac{4}{3}d_{d}^{e}-\frac{1}{3}d_{u}^{e}\approx{\cal O}\left(d_{u}^{e}\right)$
is in agreement with experiments without requiring a strong fine-tuning
of the parameter of the model~\cite{Montero2011_PQ}.
\section{Reviewing the nonthermal production of axion dark matter}\label{axiondarkmatter}
For the postinflationary $f_a$ values considered here, cold dark matter in the form of axions can be produced by three different processes: the 
misalignment mechanism~\cite{turner1991inflationary}, where the axion field oscillates about the minimum of its potential, trying to decrease the 
energy after the breaking of the PQ symmetry; and the decay of one-dimensional (global strings~\cite{harari1987evolution}) and two-dimensional 
(domain walls~\cite{vilenkin1982cosmic}) topological defects, which appear after breaking 
this symmetry. Now, we will briefly review the general expressions for the axion relic density in these three mechanisms following 
Ref.~\cite{kawasaki2015axion}.
\subsection{Misalignment mechanism}
The equation of motion for the axion field $a$ in a homogeneous and isotropic Universe,
is of the type of a damped harmonic oscillator with a natural frequency equal to the axion mass. In this case, taking into account nonperturbative effects of QCD at finite temperature and considering the interacting 
instanton liquid model (IILM)~\cite{Olivier2010}, the axion mass depends on the temperature as~\cite{wantz2010axion}
\begin{equation}\label{eq:ax-mass}
m_a^2(T)=c_T\frac{\Lambda_{\rm QCD}^4}{f_a^2}\left(\frac{T}{\Lambda_{\rm QCD}}\right)^{-n},
\end{equation}
where the values of the parameters are $c_T=1.68\times 10^{-7}$, $n=6.68$ and $\Lambda_{\rm QCD}=400\ \rm MeV$~\cite{wantz2010axion}. 
This dependence, is valid in the regime where the axion mass at temperature $T$ is less than its value at temperature zero, given by $m_a(0)^2=c_0\frac{\Lambda_{\rm QCD}^4}{f_a^2},$ where 
$c_0=1.46\times 10^{-3}$, which leads to a minimum temperature $\sim 103\ \rm MeV$ for the validity of the fit. 
The temperature $T_{\rm osc}$ at which the axion field begins to oscillate is given by~\cite{kawasaki2015axion}   
\begin{equation}\label{eq:temposc}
T_{\textrm{osc}}=2.29\ \textrm{GeV}\left(\frac{g_*(T_{\textrm{osc}})}{80}\right)^{-\frac{1}{4+n}}\left(\frac{f_a}{10^{10}\,\rm{GeV}}\right)^{-\frac{2}{4+n}}\left(\frac{\Lambda_{\rm QCD}}{400\ \rm{MeV}}\right),
\end{equation}
where $g_*(T_{\rm osc})$ is the number of relativistic degrees of freedom at temperature $T_{\rm osc}$. Eq.~(\ref{eq:temposc}) is valid for 
temperatures greater than $103\ \rm MeV$, where Eq.~(\ref{eq:ax-mass}) holds, and it is also assumed a not too strong dependence on the temperature 
of $g_*$, which, for the range 
$10^{9}\,\textrm{GeV}<f_a<10^{12}\,\textrm{GeV}$ analyzed in this work, varies between 80 and 
85~\cite{kolb1994early}, what would change the abundance of axion dark matter by a factor of $\approx 1.02$. 
Once the adiabatic condition is satisfied, both the entropy and the number of axions with 
momentum zero  per comoving volume are conserved ~\cite{PRESKILL1983127}, and it is possible to obtain the dark matter abundance ~\cite{kawasaki2015axion}
\begin{equation}\label{eq:omegamis}
\Omega_{a,\rm mis}h^2=4.63\times 10^{-3}\left(\frac{f_a}{10^{10}\rm GeV}\right)^{\frac{6+n}{4+n}},
\end{equation}
where $g_*(T_{\rm osc})=80$ and $\Lambda_{\rm QCD}=400\ \rm MeV$ have been used.\\

\subsection{Decay of global strings}
Global strings are the first of the topological defects that appear after the breaking of the U(1)$_{\rm PQ}$ symmetry at $T\lesssim v_{\phi}$ because the field $\phi$ (with PQ charge equal to $6$ in the $3-3-1$ model considered here) acquires a VEV $|\langle\phi\rangle|=v_{\phi}$ ~\cite{vilenkin1985cosmic,kawasaki2015axion}.  Actually, the 
breaking of the PQ symmetry leads to the formation of a densely knotted network of cosmic axion strings, which oscillate under their own tension, losing their energy by radiating axions~\cite{davis1986cosmic}. The radiation 
process lasts from the PQ-symmetry breaking time to the QCD phase transition time. Using results of numerical studies which provide the time dependence of $\rho_{\rm string}$ (energy density of strings) and $\rho_{a,\rm string}$ (energy density of axions produced
by the string decays), it is
possible to obtain the nowadays abundance of radiated axions~\cite{hiramatsu2011improved,hiramatsu2012production},
\begin{equation}\label{eq:stringabundance}
\Omega_{a,\rm string}h^2=\alpha  N_{DW}^2\times\left(\frac{f_a}{10^{10}{\rm GeV}}\right)^{\frac{6+n}{4+n}},
\end{equation} 
with $\alpha=(7.3\pm 3.9)\times 10^{-3}$,  $g_*(T_{\rm osc})=80$ and $\Lambda_{\rm QCD}=400\ \rm MeV$. $N_{\rm DW}=3$ is the number of domain walls in this model, and $n=6.68$ is the same parameter that appears in Eq.~(\ref{eq:ax-mass}).

\subsection{Decay of string-wall systems}
In the $3-3-1$ model considered, a $\ZZ_{3}$ subgroup remains after the breaking of the U$(1)_\textrm{PQ}$ symmetry, which 
 makes the vacuum manifold to be made of several disconnected components. When the temperature of the Universe lies
between the electroweak and QCD phase transition energy scales, domain walls appear as a consequence of breaking this $\ZZ_{3}$ discrete symmetry. 
These domain walls are attached by strings and occur at the boundaries between regions of space-time where the value of the field $\phi$ is different. These inhomogeneities of space-time are in tension with the assumptions of standard cosmology. So, it is necessary that these domain walls decay at a certain time after being formed~\cite{sikivie1982axions}. Actually, the domain walls bounded by strings begin to oscillate and eventually, when their tensions are greater than the tensions of the strings, their annihilations lead to axion production~\cite{vilenkin1994cosmic,lyth1992estimates}.

The energy density of domain walls can overclose the Universe due to its dependence on the inverse of the square of the scale factor, $R$, which 
decreases at a slower rate than the corresponding to matter, $\rho\sim R^{-3}$, and radiation, $\rho\sim R^{-4}$. In our case, this problem is solved 
by the introduction of a Planck-suppressed operator in the effective potential for the axion field $a$, parametrized as in Eq.~(\ref{gravpotential}).\\
The current axion abundance is given by the expression~\cite{kawasaki2015axion, ringwald2016axion}:
\begin{align}
\Omega_{a,\rm wall}h^2=&1.23\times10^{-6}[7.22\times 10^3]^{\frac{3}{2p}}\ \beta\left(\frac{2p-1}{3-2p}\right)\left[N_{\rm DW}^4\left(1-\cos\frac{2\pi N}{N_{\rm DW}}\right)\right]^{1-\frac{3}{2p}}\notag\\
&\times|g|^{1-\frac{3}{2p}}\left(\frac{\Xi}{10^{-52}}\right)^{1-\frac{3}{2p}}\left(\frac{f_a}{10^{10}{\rm GeV}}\right)^{4+\frac{3(4p-16-3n)}{2p(4+n)}}\label{eq:walls},
\end{align}
where $\Xi=\frac{1}{2^{\frac{N}{2}}}\left(\frac{v_{\phi}}{M_{\textrm{Pl}}}\right)^{N-4}$, and $\beta=1.65\pm 0.47$ is a parameter obtained from numerical
simulations. Finally, we will refer to the case $p=1$ as the exact scaling, and $p\neq 1$ as the 
deviation from scaling. From here on, we use $p=0.926$ for the deviation from scaling case, since it is the suggested value by numerical 
simulations~\cite{kawasaki2015axion}.\\

In order to conclude this section, we have seen that axions can be produced by three different non-thermal mechanisms, which leads to the 
result that the total abundance of axions in the Universe can be written as the sum of all these contributions, Eqs. (\ref{eq:omegamis}),
(\ref{eq:stringabundance}) and (\ref{eq:walls}), i.e.,
\begin{equation}\label{eq:totabundance}
\Omega_ah^2=\Omega_{a,{\rm mis}}h^2+\Omega_{a,{\rm string}}h^2+\Omega_{a,{\rm wall}}h^2.
\end{equation}
The total dark matter abundance due to axions is upper bounded by the observational constraint 
on the current relic density $\Omega^{\textrm{Planck}}_{\textrm{DM}}h^{2}= 0.1197\pm 0.0066$ (at $3\sigma$) as reported by the Planck Collaboration
~\cite{Ade:2015xua}. In the next section, we will analyze the behavior of each contribution to the total abundance, in order to establish a suitable 
region of parameters for the model analyzed in this work.

\section{Constraining the nonthermal production of axion dark matter}\label{results}
In general, the total dark matter relic density due to axions in this $3-3-1$ model depends on $f_a,\,g,\,N_{\textrm{DW}}$ and $\ZZ_N$. 
The dependence on $f_a,\,g,$ and $N_{\textrm{DW}}$ is direct because $\Omega_{a,{\rm mis}},\,\Omega_{a,{\rm string}}$ and $\Omega_{a,{\rm wall}}$ 
explicitly depend on these parameters. Nevertheless, the dependence on $\ZZ_N$ is indirect. Roughly speaking, this discrete symmetry constrains the 
order of the dominant gravity-induced operator $g\phi^{N}/M_{\text{Pl}}^{N-4}$. In other words, the discrete symmetry sets the exponent $N$ which 
directly affects the total dark matter due to axions. Actually, we have two discrete symmetries, $\mathbb{Z}_{10}$ and $\mathbb{Z}_{11}$ 
(see Table \ref{table 3}), that stabilize the PQ mechanism, which implies that there are two cases to be considered, $N=10$ and $N=11$. On the other 
hand, the domain wall parameter, $N_{\textrm{DW}}$, is set to be equal to $3$ by the PQ symmetry and the matter content in the model. Thus, we 
are interested in knowing if the model with $\mathbb{Z}_{10}$ or/and $\mathbb{Z}_{11}$ symmetry provides the total dark matter reported by the 
Planck collaboration~\cite{Ade:2015xua} when $f_a,\,g,$ take their allowed values, without conflicting with the constraints on the axion phenomenology.

\begin{figure}[!htb]
\begin{center}
 \includegraphics[scale=1]{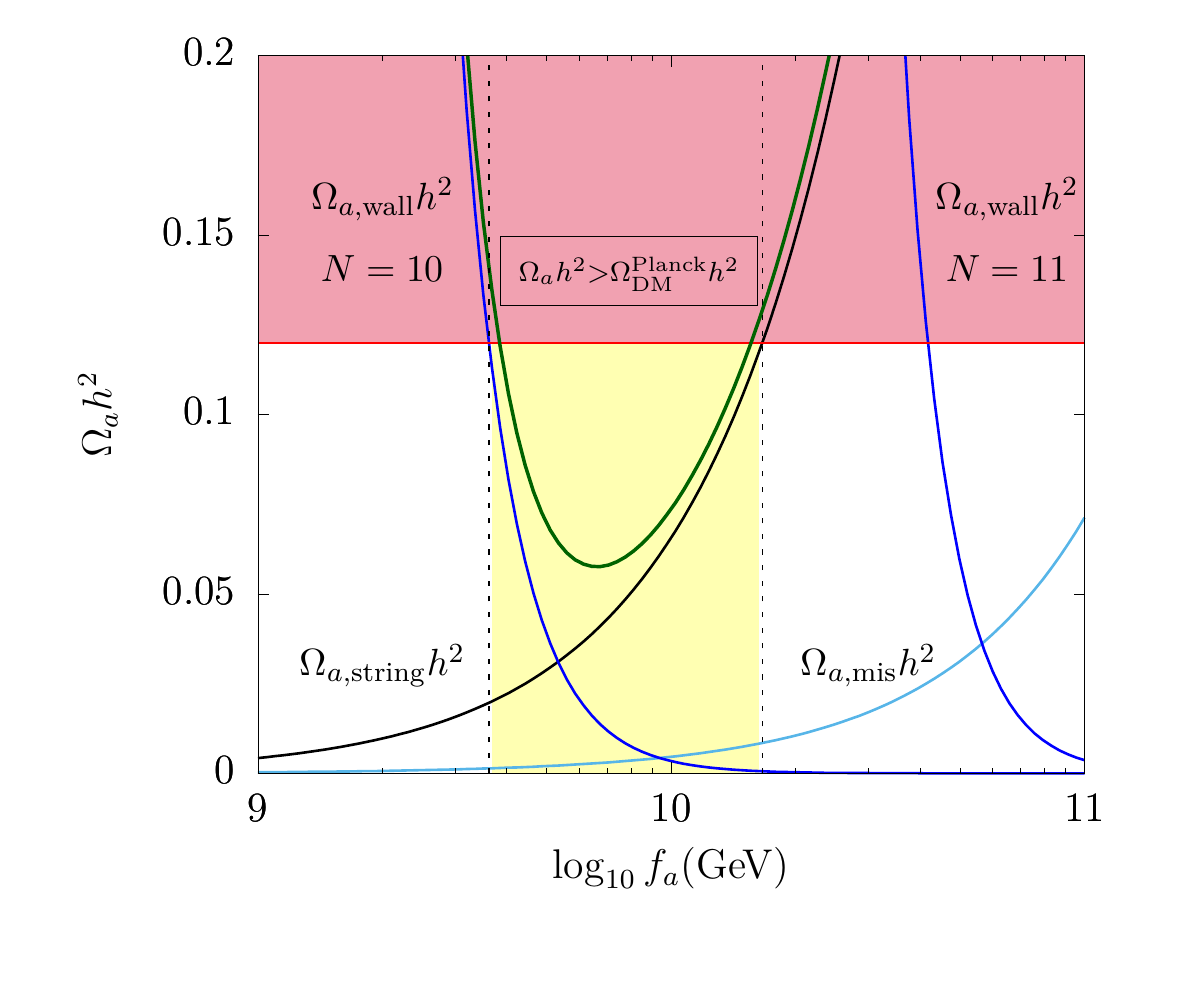}
 \caption{Relic density of nonthermal axion dark matter in the $3-3-1$ model, assuming exact scaling, $p=1$, and $|g|=1$. The central values of the parameters in Eqs. (\ref{eq:stringabundance}) and (\ref{eq:walls}) together with $N_{\textrm{DW}}=3$ have been used. The vertical dashed lines limit regions with over production of axions by decay of domain walls (left line) and strings (right line), while the horizontal red line is the experimental constraint $\Omega_ah^2=\Omega^{\textrm{Planck}}_{\textrm{DM}}h^{2}$.\label{Abundancias}}
\end{center}
\end{figure}

In order to do that, it is convenient, first, to study separately the behavior of the three axion production mechanisms which results are shown in Fig.~\ref{Abundancias}. Specifically, the cyan and black lines show the axion abundances produced by misalignment and global string decay mechanisms, respectively. On the other hand, the blue lines show the abundance of axion dark matter due to the decay of domain wall systems for $N=10$ and $N=11$, calculated for the coupling constant value $|g|=1$. Two shaded regions are also shown: the light red one corresponds to the exclusion region coming from the constraint of the over closure of the Universe~\cite{Ade:2015xua}, and the yellow region gives the possible interval for the axion decay constant $f_a$, for which no over abundance of axions from decay of global strings or domain walls is produced. Finally, the dark green line corresponds to the total abundance of axions, $\Omega_ah^2$, as given by Eq.~(\ref{eq:totabundance}), obtained for the case $N=10$ and $|g|=1$. The case for $N=11$ is not shown because for all the considered values of $f_a$ the axion relic density is overabundant. 

\begin{figure}[!htb]
	\centering
 	\begin{subfigure}[t]{0.47\textwidth}
		\centering
		\includegraphics[scale=0.7]{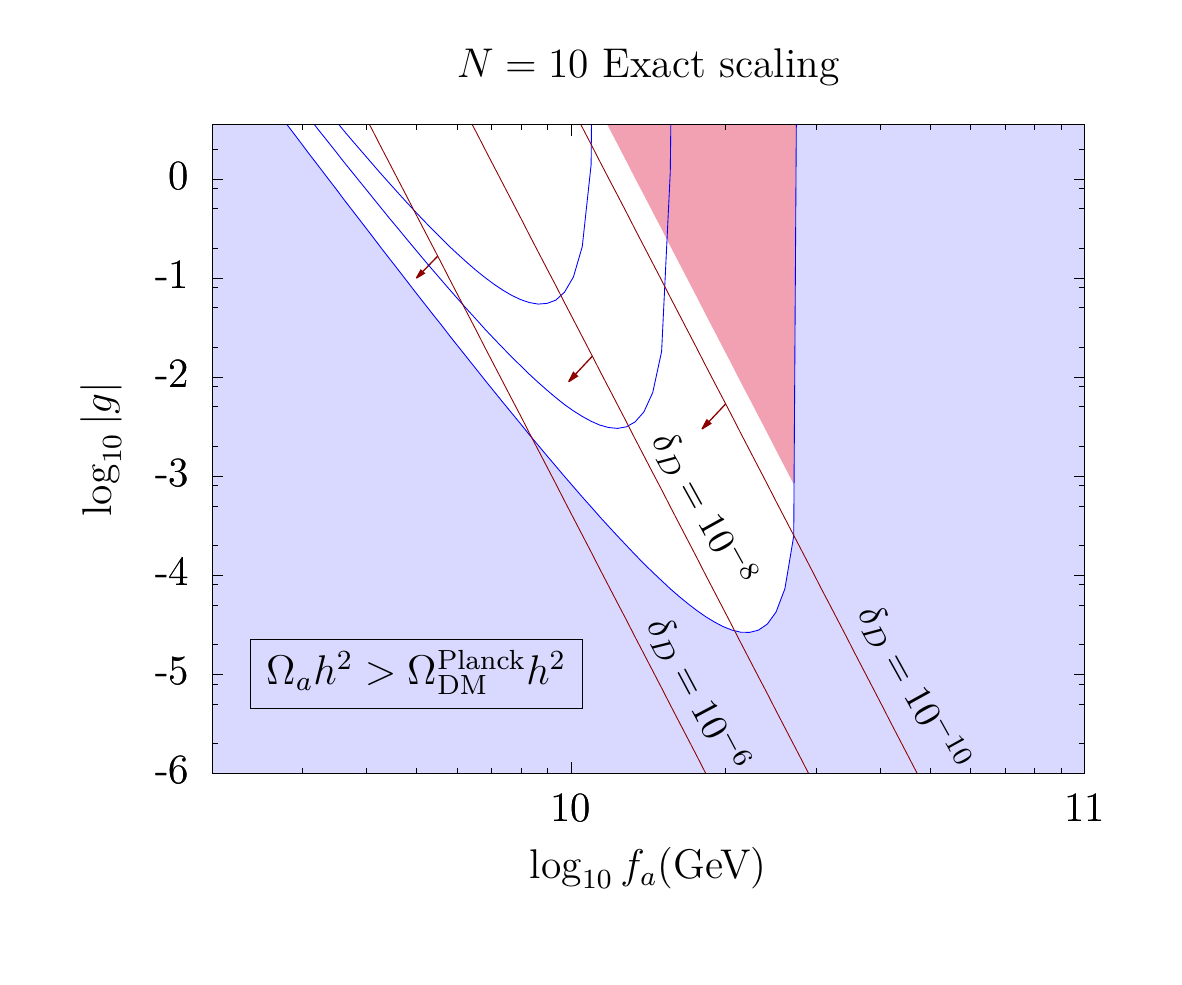}
    \caption{} \label{fig:fig_a}
	\end{subfigure}%
	\hfill
	\begin{subfigure}[t]{0.47\textwidth}
		\centering
		\includegraphics[scale=0.7]{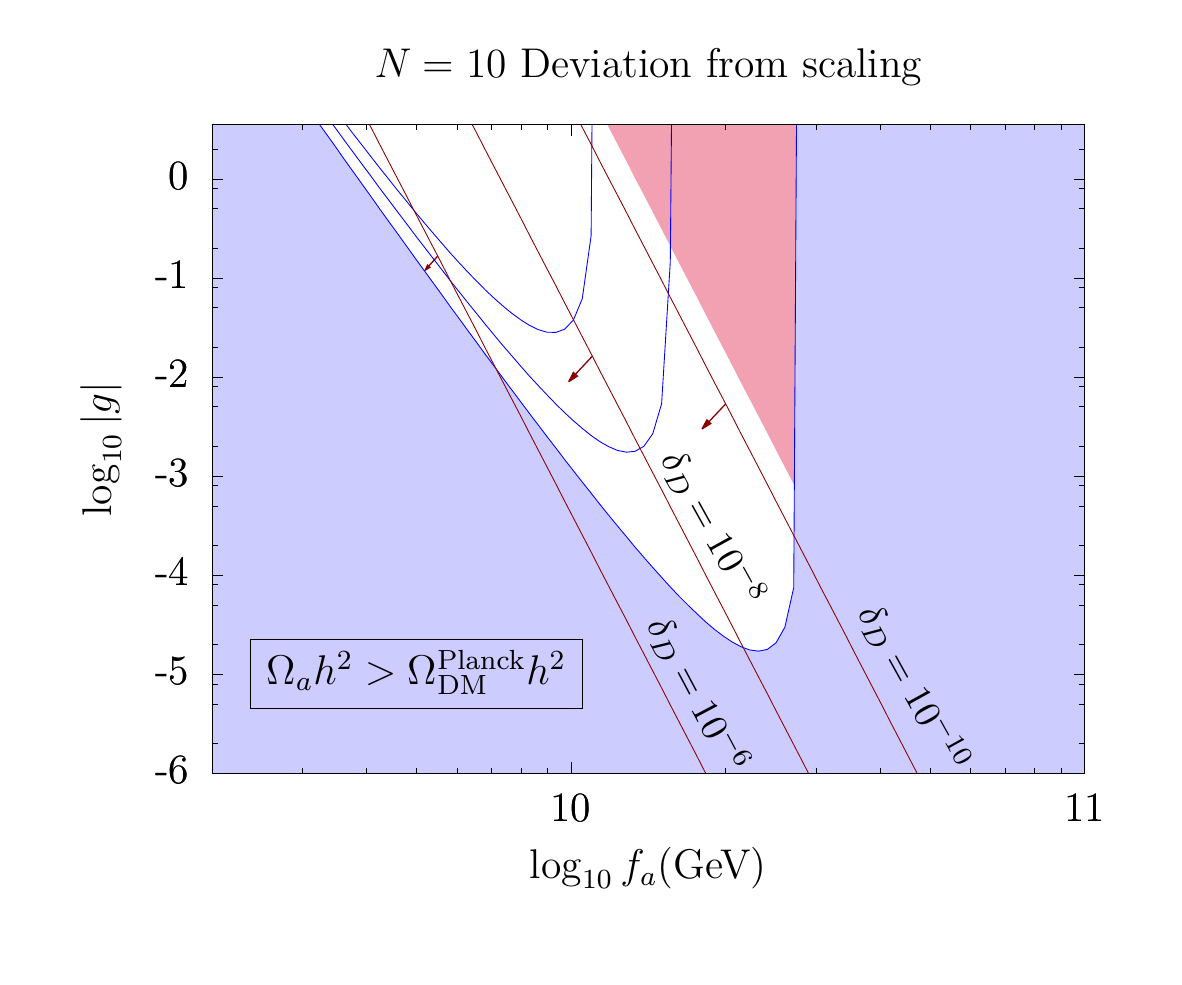}
		\caption{} \label{fig:fig_b}
	\end{subfigure}%
\caption{Observational constraints on the parameter space $f_a-|g|$ in the $3-3-1$ model, assuming exact scaling (a) and deviation from scaling (b). 
These plots correspond to  the $\mathbb{Z}_{10}$ discrete symmetry, and $N_{DW}=3$. The shaded regions in light red and light blue correspond to 
regions of the parameter space where the constraints given by $m_{a,{\rm QCD}}>m_{a,{\rm{gravity}}}$ and 
$\Omega_{a}h^2\leqslant\Omega^{\textrm{Planck}}_{\textrm{DM}}h^{2}$ are violated, respectively. Moreover, the regions above the straight red lines 
correspond to the exclusion regions set by the NEDM condition, as given by Eqs. (\ref{labelNEDM1}) and (\ref{eq:NEDM}), for three different choices 
of the  $\delta_D$ parameter. }~\label{EspacioParametrosN10}
\end{figure}

From Fig.~\ref{Abundancias} some conclusions are straightforward. First, $\Omega_{a,{\rm mis}}$ and $\Omega_{a,{\rm string}}$ grow when $f_a$ grows. 
Thus, in principle, these are dominant for the greater values of $f_a$ ($5.3\times 10^9\, {\rm{GeV}}\lesssim f_{a} \lesssim 1.7\times10^{10}\ 
\rm{GeV}$). However, the misalignment mechanism is always subdominant because $\Omega_{a,{\rm string}}$ has an extra $N^{2}_{\textrm{DW}}=9$ global 
factor. Indeed, the misalignment mechanism contributes at most by $\approx7\%$ for the total dark matter density. In contrast,
$\Omega_{a,{\rm wall}}$ is decreasing with $f_a$ and thus it dominates $\Omega_a$ for the smaller values of
$f_a$ ($3.6\times10^9\ {\rm GeV}\lesssim f_a\lesssim 5.3\times 10^9\ \rm{GeV}$). That can be understood realizing that the domain-wall time
decay is larger for smaller $f_a$ values, making the domain wall more stable and, in this way, explaining why this mechanism contributes more  
for the axion relic density when $f_a$ is smaller. The opposite behavior of $\Omega_{a,{\rm string}}$ and  $\Omega_{a,{\rm wall}}$ allow to set an
upper and lower bound on $f_a$. For $|g|=1$, $f_a$ is constrained to be $3.6\times 10^9\ {\rm GeV}<f_a<1.7\times10^{10}\ \rm{GeV}$ in order to 
satisfy $\Omega_{a,{\rm wall}}h^2\,\textrm{ and }\Omega_{a,\rm string}h^2\lesssim \Omega^{\textrm{Planck}}_{\textrm{DM}}h^{2}$~\cite{Ade:2015xua}.
Actually, the interval of allowed $f_a$ values is slightly thinner because all of the three axion production mechanisms contribute simultaneously.
Also, note that the $f_a$ upper bound above is independent on the value of $N$ and on the value of $|g|$, as can be seen from 
Eq.~(\ref{eq:stringabundance}). In contrast, the lower bound is only valid for the case of $N=10$. Actually, the case of $\mathbb{Z}_{11}$ is 
completely ruled out and, for this reason, our analysis will be concerned exclusively with the $\ZZ_{10}$ symmetry case. 
Once we have gained a general knowledge about the behavior of $\Omega_ah^2$ as function of $f_a$ for $|g|=1$, we can go further studying the
parameter space for the $\mathbb{Z}_{10}$ case, allowed by the axion phenomenology. In particular, in Fig.~\ref{EspacioParametrosN10} we show 
the parameter space $f_{a}-|g|$ for the cases of exact scaling ($p=1$, left frame) and deviation from scaling ($p=0.926$, right frame). 
The range of values of the coupling constant, $g$, has been chosen to include values of $|g|\leq\sqrt{4\pi}$. The blue curves correspond to the 
regions where the total axion dark matter abundance is equal to $\Omega^{\textrm{Planck}}_{\textrm{DM}}h^{2}$, taking into account the uncertainties
in the parameters $\alpha$  and $\beta$ in Eqs. (\ref{eq:stringabundance}) and (\ref{eq:walls}). Notice that for a given value of $f_a$, $|g|$ is 
lower bounded by these lines.  Larger values of $|g|$ imply $\Omega_{a}h^2<\Omega^{\textrm{Planck}}_{\textrm{DM}}h^{2}$. The light blue shaded 
region is ruled out by the over closure of the Universe for the case of the parameter $\beta=2.12$ in Eq. (\ref{eq:walls}) and for 
the $\alpha=7.3-3.9=3.4$ factor in Eq. (\ref{eq:stringabundance}). From the remaining region, it is possible to exclude another large part 
applying the axion mass stability condition, $m_{a,{\rm QCD}}>m_{a,{\rm{gravity}}}$ [see the discussion near Eq. (\ref{eq:mgravity})]. 
Because $m_{a,\textrm{ gravity}}$ is directly proportional to $|g|$ and $f_a^{N-2}$, cf. Eq.~(\ref{eq:mgravity}), and $m_{a,\textrm{ QCD}}^{2}$ 
is inversely proportional to $f_a^{2}$, cf. Eq. (\ref{eq:axionQCDmass}), the forbidden region, denoted by the light red color, is in the top right
part of the $f_{a}-|g|$ plane. In addition, in Fig.~\ref{EspacioParametrosN10} are shown three  dark red lines which correspond to the NEDM 
constraint, given by Eq.~(\ref{eq:NEDM}), for different values of $\delta_D$. It is important to realize that $\delta_D$ values of order one 
(not shown) do not give allowed regions in the parameter space. It is necessary to allow $\delta_D\lesssim 10^{-5}$ in order to have nonexcluded 
regions which are below the lines. In particular, we calculate the maximum values of $\delta_D$ that give allowed regions in the parameter space. 
The corresponding results, in the cases of exact scaling ($p=1$) and deviation from scaling ($p=0.926$), are
\begin{equation}
\delta_{D}=\begin{cases}
(0.4-4.1)\times 10^{-5} & \textrm{Exact scaling},\\
(2.9-9.5)\times10^{-6} & \textrm{Deviation from scaling.}
\end{cases}
\end{equation}
These values are obtained by taking $|g|=\sqrt{4\pi}$, and considering the uncertainties in the parameters of the three  axion production mechanisms. Lower values of $|g|$ would require higher tuning on the $\delta_D$ parameter, with values of the order $10^{-8}$ as shown in Fig.~\ref{EspacioParametrosN10}. In general, for $|g|$ fixed, the tuning on $\delta_D$ depends on the decay constant $f_a$ and the mechanism of axion dark matter production: if the decay of domain walls was dominant (left side of the curves), the  tuning would be less severe than if the production by string decay (right side of the curves) was the dominant one.

Also, in Fig.~\ref{EspacioParametrosN10} is shown that for a $\delta_D$ small enough in order to satisfy the NEDM condition, and for a given $|g|$ value between $5\times10^{-2}$ and $\sqrt{4\pi}$, there are two separated regions for $f_a$ where axions can make up the total DM relic density. For instance, taking $|g|=\sqrt{4\pi}$ and considering the uncertainties in the parameters, these regions and their corresponding axion masses for the exact scaling case, are
\begin{equation}\label{eq38}
 f_a\approx \begin{cases*}
             (2.8-3.5)\times 10^9\ {\rm GeV}\quad\,\,\,\longrightarrow \quad  m_a \approx (1.7-2.1)\times 10^{-3}\ \rm{eV} \\
             (1.1-1.2)\times10^{10}\ {\rm GeV} \quad \longrightarrow \quad  m_a\approx(5-5.4)\times 10^{-4}\ \rm{eV}
            \end{cases*}
\end{equation}
In the first range for $f_a$ the production of dark matter is mainly through the decay of domain walls, while in the second range it is due to the decay of strings. Taking smaller values for $|g|$, will lead to more stringent
intervals for both $f_a$ and $m_a$. For the case of deviation from scaling, we find $f_a\approx (3.4-3.6)\times10^9\ {\rm GeV}$, corresponding to 
$m_a\approx(1.7-1.8)\times 10^{-3}\ {\rm eV}$, when the domain walls decay is the leading production mechanism, and 
$f_a\approx(1.1-1.2)\times10^{10}\ {\rm GeV}$, leading to $m_a\approx (5-5.4)\times 10^{-4}{\,\rm{ eV}}$, for the  string decay as the dominant contribution.

\begin{figure}[!htb]
		\resizebox{0.7\textwidth}{!}{\includegraphics{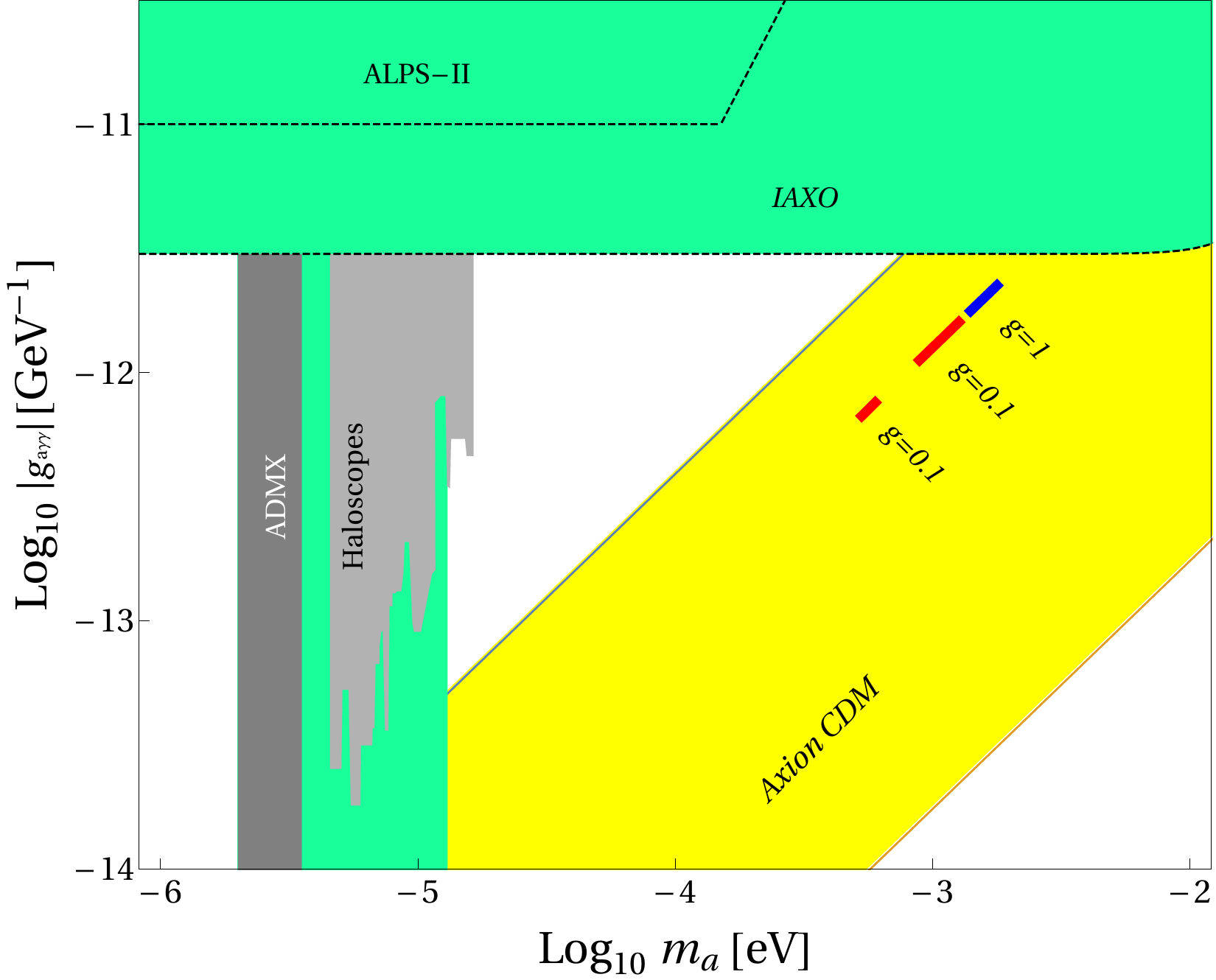}}
\caption{Projected sensitivities of different experiments in the search for axion dark matter. The green regions show sensitivities of light-shining-through-wall experiments like ALPS-II~\cite{ALPSII}, of the helioscope IAXO~\cite{DAFNI2016244}, of the haloscopes ADMX and ADMX-HF~\cite{Bahre:2013ywa,Armengaud:2014gea}. The yellow band corresponds to the generic prediction for axion models in QCD. In addition, the two (one) thick red (blue) lines stand for the predicted mass ranges and coupling to photons in this model, for $|g|=0.1$ ($|g|=1$), where axions make up the total DM relic density.\label{Parameterspace}}
\end{figure}

Finally, for values of $\vert g \vert$ of order one, we can make predictions regarding the observability of axion in current and/or future experiments. Specifically, the axion coupling to two photons, $g_{a\gamma\gamma}$, depends on the $f_a$ decay constant, the electromagnetic and color anomaly coefficients, $E$ and $N_C$, respectively. It is known that these anomaly coefficients are completely determined by the fermion content and the U$(1)_{\textrm{PQ}}$ charges of the model, cf. Table~(\ref{table 2}). Standard calculations for anomaly coefficients~\cite{Srednicki:1985xd,Carvajal:2015dxa} furnish $E=-4$ and $N_C=3$. With this information, we can go further plotting, in Fig~\ref{Parameterspace}, $g_{a\gamma\gamma}$ as a function of $m_a$ for the regions where axions make up the total dark matter relic density and for two different values of $|g|$, specifically $|g|=0.1$ and $|g|=1$. This figure clearly shows two allowed regions for $|g|=0.1$: $m_a\approx(0.4-0.6)\times 10^{-3}\ \rm{eV}$ with $g_{a\gamma\gamma}\approx(4.5-5.9)\times 10^{-13}\ \rm{GeV}^{-1}$ and $m_a \approx (0.9-1.3)\times 10^{-3}\ \rm{eV}$ with $g_{a\gamma\gamma}\approx(1.1-1.6)\times 10^{-12}\ \rm{GeV}^{-1}$, and one region for $|g|=1$: $m_a\approx(1.4-1.8)\times 10^{-3}\ \rm{eV}$ with $g_{a\gamma\gamma}\approx(1.8-2.2)\times 10^{-12}\ \rm{GeV}^{-1}$. The reason why there is only one region for larger $|g|$ values is that the gravitational mass grows with $|g|$ and thus, it conflicts with the condition $m_{a,\textrm{ QCD}}\gg m_{a,\textrm{ gravity}}$ for lower axion masses. Moreover, it is notable that for the range with larger masses (blue line), the axion parameters of this $3-3-1$ model are very close to the projected region which is going to be explored by the IAXO 
experiment~\cite{ringwald2016axion,DAFNI2016244}.
	
\section{Conclusions\label{conclusions}}
In this work, we consider a version of an alternative electroweak model based on the
$\textrm{SU}(3)_L\otimes \textrm{U}(1)_X$ gauge symmetry, the so called $3-3-1$ models, when the color gauge group is added. For this 
version, which includes right-handed neutrinos, it is shown in Ref.~\cite{Montero2011_PQ} that the PQ mechanism for the solution of the strong $CP$ problem can be implemented. In this implementation, the axion, the pseudo Nambu-Goldstone boson that emerges from the PQ-symmetry breaking, is made invisible by the introduction of the scalar singlet $\phi\sim (1,1,1)$ whose VEV, $v_\phi \approx {\tilde{f}_a}$, is much larger than  $v_{\textrm{SM}}$, and any other VEV in the model. Moreover, the axion is also protected against gravitational effects, that could destabilize its mass, by a discrete $\mathbb{Z}_{N}$ symmetry, with $N=10,11$.  

Once we have set this consistent scenario, we investigate the capabilities of this axion, produced in the framework of this particular  $3-3-1$ model,
to be a postinflationary cold dark matter candidate. We started focusing in the axion-production mechanisms. As it was explained in the previous section, from Fig.~\ref{Abundancias} we see that the vacuum misalignment mechanism does not dominate the DM relic abundance, and, if it was the only production mechanism in action, an upper bound for $f_a$ could be set by imposing that it should account for all the  DM abundance, i.e., $\Omega_{a,\textrm{mis}}h^2= \Omega_\textrm{DM}^\textrm{Planck}h^2$, and we would find the corresponding value $f_a \approx 1.5 \times 10^{11}\,\textrm{GeV}$, for the parameters determined by the model, in this case $N_{DW}=3.$ However, there are two other more efficient mechanisms  due to the decay of topological defects: cosmic strings and domain walls. As the curves for $\Omega_{a,\textrm{string}}h^2$ and $\Omega_{a,\textrm{wall}}h^2$ grow in opposite directions, relatively to the $f_a$ values, we can determine an upper bound and a lower bound for $f_a$ by imposing  the total $\Omega_ah^2$ matches the observed Planck results. This is the case when we add up all the contributions for $N=10$, and we find $3.6 \times 10^9\,\textrm{GeV} < f_a < 1.7 \times 10^{10}\,\textrm{GeV}$. However, we would like to stress that this is not the case for $N=11$. For $N=11$ there is no value of $f_a$ for which the addition of the partial abundances lies below the observed result. It means that the  $\mathbb{Z}_{11}$, which possesses the good quality of stabilizing the axion, is not appropriate for the axion-production issue since it makes the domain wall mechanism too efficient and overpopulates the Universe. 

As it can be seen from Fig.~\ref{Abundancias}, for any fixed allowed value of $\vert g \vert$, there are two values of $f_a$ that are in agreement with the value of $\Omega_\textrm{DM}^\textrm{Planck}h^2$. In fact they are regions, if we take into account the uncertainties following the discussion in the previous section for Fig.~\ref{EspacioParametrosN10}. 
Outside these regions, the axion abundance will be a fraction of $\Omega_\textrm{DM}^\textrm{Planck}h^2$.
See the solid dark green curve in Fig.~\ref{Abundancias} for $\vert g \vert =1$. If this happens to be the case, i.e., if these predicted regions 
are somehow excluded, by future experimental data for the axion mass value, for instance, then, another kind of DM will be needed. We have also 
found special values for $\delta_D$, $(0.4-4.1)\times 10^{-5}$, by requiring the minimal compatible intersection region between the curves that obey 
the NEDM and $\Omega_\textrm{DM}^\textrm{Planck}h^2$ constraints. This value was obtained considering the maximum value of $\vert g \vert$, i.e.,
$\vert g \vert = \sqrt{4\pi}$, cf. Fig.~\ref{EspacioParametrosN10}\subref{fig:fig_a}. For lower values of $\vert g \vert$, higher tuning on $\delta_D$ is required. However, it
seems unnatural to require severe levels of tuning on $\delta_D$, since for this quantity a {\it tiny} value is the result of the difference between 
two terms that have completely different origins.  

Regarding the capabilities of detecting the axion dark matter, Fig.~\ref{Parameterspace} shows the sensitivities of several experiments in 
the $m_a-g_{a\gamma\gamma}$ plane.  In this plot, the  thick blue and red lines are the regions where the axion abundance is responsible for all 
the observed DM. These lines were obtained by using $\vert g \vert$ of order one. Moreover, the blue region corresponding to masses of the order 
of meV and $g_{a\gamma\gamma}\approx 10^{-12}\,\textrm{GeV}^{-1}$, lies very close to the projected IAXO sensitivity, so that it will be reachable
in the near future. 

Looking back to our results we can conclude that this version of the $3-3-1$ model, concerning the axion DM issue and the strong $CP$ problem, is phenomenologically consistent. This model,  besides its good qualities presented in the introduction,  also possesses new degrees of freedom that are not yet experimentally probed. For instance, the model has charged and neutral scalars (besides the Higgs), extra vector bosons and extra quarks, that are expected to be heavy, and could, in principle, be searched at colliders. See Refs.~\cite{Cao:2016uur,Corcella:2017dns} for recent studies concerning the $3-3-1$ model phenomenology, in general, at the LHC.

\begin{acknowledgments}
B. L. S. V. is thankful for the support of FAPESP funding Grant No. 2014/19164-6. A. R. R. C would like to thank Coordena\c{c}\~ao de Aperfei\c{c}oamento de Pessoal de N\'ivel Superior (CAPES), Brasil, for financial support. 
\end{acknowledgments}

\bibliography{referencesfv}

\end{document}